\documentclass[sigconf]{acmart}

\usepackage{graphicx} 
\usepackage{natbib}
\usepackage{booktabs}
\usepackage{tabularray}
\usepackage{multirow}
\usepackage{adjustbox}
\usepackage{amsmath}
\usepackage{comment}
\usepackage{subcaption}
\usepackage{hyperref}
\usepackage{arydshln}
\usepackage{xspace}

\usepackage[utf8]{inputenc}
\usepackage{pgfplots}
\DeclareUnicodeCharacter{2212}{−}
\usepgfplotslibrary{groupplots,dateplot}
\usetikzlibrary{patterns,shapes.arrows}
\pgfplotsset{compat=newest}

\newcommand{\noncons}[1][]{$NU$\xspace}
\newcommand{\consmp}[1][]{U$_{aw}$\xspace}
\newcommand{\consnomp}[1][]{U$_{ag}$\xspace}

\newcommand{\paragraphHd}[1] {\vspace{3pt}\noindent\textbf{#1}}

\AtBeginDocument{%
  }

\setcopyright{acmlicensed}
\copyrightyear{2018}
\acmYear{2018}
\acmDOI{XXXXXXX.XXXXXXX}

\acmConference[Conference acronym 'XX]{Make sure to enter the correct
  conference title from your rights confirmation emai}{June 03--05,
  2018}{Woodstock, NY}
\acmISBN{978-1-4503-XXXX-X/18/06}

\begin{document}

\title{Transfer Learning for E-commerce Query Product Type Prediction}

\author{Anna Tigunova}
\affiliation{%
  \institution{Amazon}
  \city{Berlin}
  \country{Germany}}
\email{tigunova@amazon.com}

\author{Thomas Ricatte}
\affiliation{%
  \institution{Amazon}
  \city{Luxembourg}
  \country{Luxembourg}}
\email{tricatte@amazon.com}

\author{Ghadir Eraisha}
\affiliation{%
  \institution{Amazon}
  \city{Luxembourg}
  \country{Luxembourg}}
\email{eraishag@amazon.com}

\renewcommand{\shortauthors}{Trovato et al.}

\begin{abstract}

Getting a good understanding of the customer intent is essential in e-commerce search engines. In particular, associating the correct \emph{product type} to a search query plays a vital role in surfacing correct products to the customers. 

Query product type classification (Q2PT) is a particularly challenging task because search queries are short and ambiguous, the number of existing product categories is extremely large, spanning thousands of values. Moreover, international marketplaces face additional challenges, such as language and dialect diversity and cultural differences, influencing the interpretation of the query.

In this work we focus on Q2PT prediction in the global multi-locale e-commerce markets. The common approach of training Q2PT models for each locale separately shows significant performance drops in low-resource stores. Moreover, this method does not allow for a smooth expansion to a new country, requiring to collect the data and train a new locale-specific Q2PT model from~scratch.

To tackle this, we propose to use \textit{transfer learning} from the high-resource to the low-resource locales, to achieve global parity of Q2PT performance. We benchmark the per-locale Q2PT model against the \textit{unified} one, which shares the training data and model structure across all worldwide stores. Additionally, we compare locale-aware and locale-agnostic Q2PT models, showing the task dependency on the country-specific traits.
%This design allows to transfer knowledge from the data-rich locales, alleviating the cold-start problem for Q2PT prediction in new stores. 
%To assess if the Q2PT task is locale-invariant we compare locale-aware and -agnostic Q2PT models, showing that the model provided with the information about the customer's locale performs better, and that there are numerous cases where Q2PT prediction for a search query varies depending on the store.

We conduct extensive quantiative and qualitative analysis of Q2PT models on the large-scale e-commerce dataset across 20 worldwide locales, which shows that unified locale-aware Q2PT model has superior performance over the alternatives. %Additionally, we study the important factors that influence Q2PT models to be sensitive to locale differences. Our insights from comparing non-unified and unified Q2PT models can be useful for the industrial and academic practitioners to design their query classification systems with locale particularities in mind.

\end{abstract}


\begin{CCSXML}
<ccs2012>
<concept>
<concept_id>10002951.10003317.10003325.10003327</concept_id>
<concept_desc>Information systems~Query intent</concept_desc>
<concept_significance>500</concept_significance>
</concept>
</ccs2012>
\end{CCSXML}

\ccsdesc[500]{Information systems~Query intent}

\keywords{Query Understanding, Product Search}

\received{20 February 2007}
\received[revised]{12 March 2009}
\received[accepted]{5 June 2009}

\maketitle

\section{Introduction}
\label{sec:intro}

\paragraphHd{Problem statement.} Query understanding in e-commerce extracts customer shopping intent from their search queries, by classifying the query as having a target brand, color, size, etc. The extracted attributes are extremely important for search results ranking and filtering, query augmentation, recommendations and many other usecases \cite{chang2020query}. One of the most critical components in query understanding isa Query-to-Product Type (Q2PT) classifier, which associates customer search query with a product type (PT) that the customer intended to browse. 

Q2PT signal has a direct impact on the customer experience, as it can significantly alter search results shown to the user.
For example, for a search query ``\textit{harry potter mug}'', even if the matched Harry Potter \textit{book} has a high TF-IDF score, it should not be surfaced in the search results, since the customer is interested in another product type (\textit{mugs}). Moreover, Q2PT signal helps to optimize the computation of the search results, narrowing the retrieval to a specific product category shard \cite{lin2020light}, which yields search latency improvements.

Query category prediction has received significant attention in related works \cite{liu2019system, jiang2022short, zhu2022enhanced, zhang2021modeling, zhu2023hcl4qc}: the proposed approaches target various challenges associated with Q2PT prediction, such as short queries~\cite{liu2019system, jiang2022short}, long-tail queries \cite{zhu2022enhanced, zhang2021modeling}, product type hierarchy~\cite{zhu2023hcl4qc}, etc. However, to the best of our knowledge, there has not yet been studies, which target the issues of Q2PT classification in the multi-locale setting.

In international marketplaces, query understanding and ranking models are often trained on a per-locale basis, using the data from a single store \cite{luo2021q2pt, bonab2021cross}. This approach ensures that the peculiarities of various locales are captured: for instance, in case of Q2PT prediction, same keywords may convey different product type intents, depending on the store (query `pants' in UK would mean that the user is searching for \textit{underwear}, while in the US it means the intent to buy \textit{trousers}); \citet{bonab2021cross} show that users in multiple locales have different preferences for the same product set.

Most related studies develop and test query understanding models for high-resource locales, such as United States, achieving remarkable results.
However, many recently launched stores in new countries suffer from data shortage \cite{ferwerda2016exploring}, which is a major blocker for applying per-locale Q2PT models. 

\begin{figure}[t!]
        \centering
        \includegraphics[width=0.95\linewidth]{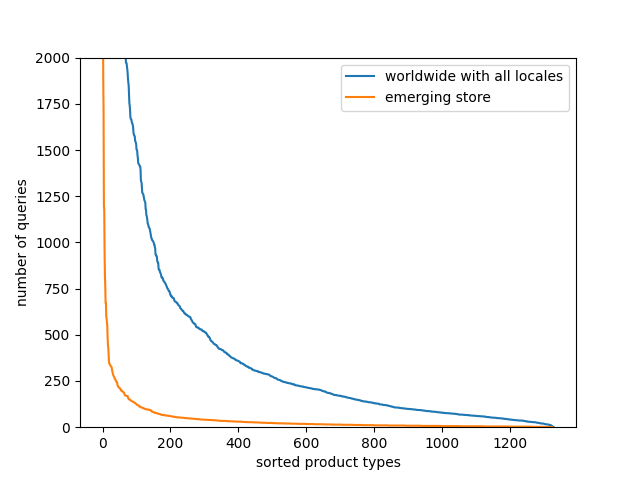}
    \caption{Query distribution for product types worldwide versus an emerging store on an experimental data sample including 20 locales and 1414 product types.}
        \label{fig:PT_percent}
    \vspace{-1.6cm}
\end{figure}

The issue is further aggravated by the long-tail nature of the product category distribution; in large online stores there are thousands of product types, with a small fraction of most popular categories dominating the distribution \cite{zhang2019improving}. This discrepancy is amplified in the emerging stores, as shown in Figure \ref{fig:PT_percent}, where we use our experimental dataset to plot the number of queries per product type, sorted by their frequency, in the emerging locale and aggregated worldwide. In both graphs, we observe that the majority of queries are covered with a couple of most popular categories, which increases the risk of the classifier overfitting and underperforming on the long-tail product types.

A science challenge in e-commerce is the cold-start problem when expanding to a new locale; to cater for this case it is important to develop query understanding models that will be capable of adapting to the new marketplces without long and costly data collection and retraining.%\anna{in Figure 1 removed axes scales, changed the caption as proposed}

\paragraphHd{Proposed approach and contributions.} To overcome the mentioned issues with sparse and unbalanced data in low-resource locales, we propose to leverage rich data in high-resource locales via \textit{knowledge transfer} for Q2PT problem.

We experiment with a unified multilingual multi-locale Q2PT model, which shares model parameters and training data across all stores. This solution allows small loacles to benefit from the knowledge of the established stores, solving the problem of data sparsity and increasing sample efficiency. Moreover, it reduces the training time and complexity, as well as memory requirements for storing a model checkpoint per each locale individually.

A possible disadvantage of this approach is that the unified \textit{locale-agnostic} model can transfer biases from the large stores to the predictions on the low-resource ones \cite{bonab2021cross}. To alleviate this, we propose a \textit{unified locale-aware} model variant, by conditioning the prediction on the locale-id. The experiments show that locale-aware model performs better than the agnostic one, which demonstrates that it can better preserve locale-specific traits.

Further, we conduct extensive qualitative analyses on the query level, to assess how much the locales' differences affect Q2PT predictions. We find that Q2PT task is not locale-invariant (i.e. the same query yields different product type distributions depending on the country) and build categorization of local differences for product type predictions.

To support our findings we conduct large-scale experiments with an experimental e-commerce data set sampled from real data, including 20 locales and 1414 product types. Our insights from comparing non-unified and unified Q2PT models will be useful for the industrial and academic practitioners, helping to design query understanding systems with locale particularities in mind.

\begin{comment}
The salient contributions of this work are as follows:
\begin{enumerate}
    \item we propose a unified Query-to-Product type model for multi-locale e-commerce marketplaces, which streamlines model training and reduces storage requirements,%simplifies and speeds up inference, while still preserving the nuances of the individual marketplaces,
    \item we conduct an extensive study comparing the per-locale and unified models, as well as locale-aware unified model,
    \item we conduct an online user study, which shows that the unified locale-aware model is prefered by the users over other alternatives.
    %\item in offline experiments we show that the proposed model has greater recall, specifically on new and emerging marketplaces, 
    %\item our online study shows that using our consolidated model in production results in increased OPS, significantly improved latency and reduction of irrelevant search results.
\end{enumerate}

%The rest of the paper is structured as follows:
\end{comment}

\section{Related Work}
\label{sec:related}

\paragraphHd{Search query classification.} E-commerce query understanding signals are essential for search rankers and other downstream services, and thus have received significant attention from the research community \cite{lin2018commerce, liu2019system, qiu2022pre, zhu2023hcl4qc, zhu2022enhanced, skinner2019commerce, zhang2021modeling, jiang2022short}. 

An important challenge in Q2PT is classifying long-tail queries, which have scarce training data. \citet{zhu2022enhanced} improve long-tail query classification by transferring knowledge from the frequent queries, which are similar to them. 
The effects of query variation are studied in \cite{zhang2021modeling}; the authors notice that slight variation to the query can flip the category prediction, and they train the classification model to distinguish between pairs of queries, which share most of the terms but have different intent. This training strategy helps to overcome data sparsity for long-tail queries. 

Multiple studies propose solutions to tailor existing classification models for short and ambiguous search queries. For instance, \citet{liu2019system} propose a hybrid system which uses different models to serve long and short queries. \citet{jiang2022short} develop a new pretraining task to improve over the standard term masking, which is detrimental for short texts: instead, they concatenate the input query with generated words and pretrain the model on identifying the extra terms. 

Another issue is the shortage of query classification training data. To tackle this issue, \citet{skinner2019commerce} propose to use transfer learning to infer query category using the model trained on product-category pairs. Alternatively \citet{qiu2022pre} propose generating synthetic queries, obtained from the product description substring, to pretrain the transformer models.

Finally, there are various other strategies proposed to improve search query classification. \citet{zhu2023hcl4qc} incorporate category contrastive loss into the model training, showing performance improvement for query classification with hierarchical classes. \citet{zhang2019improving} augment query category prediction by jointly training it with query-item semantic matching in a multi-task framework. 

\paragraphHd{Handling data from multiple locales.} Cross-market recommendations and search is an important problem for international e-commerce stores: the developed solutions need to meet the performance bar globally and be able to generalize to new locales. However, there has been only a few studies addressing this setup~\cite{roitero2020leveraging, ferwerda2016exploring, bonab2021cross}.

%\cite{zhang2021queaco} study NER in e-commerce queries across 12 languages.
\begin{comment}
    
\begin{figure*}[htp!]
    \centering
        \centering
        \includegraphics[width=0.85\linewidth]{figures/new_scheme2.pdf}
    \caption{The architecture of the baseline and consolidated labels models.}
    \label{fig:model_architecture}
\end{figure*}

\end{comment}

\citet{bonab2021cross} investigate market adaptation through fine-tuning a general model with limited data from a new locale. In this work, the product recommendation model is pretrained on all locales, but afterwards is fine-tuned on each store separately. Such approach helps the models to share some knowledge, but it complicates model training and serving, and incurs high storage costs. The method proposed in our paper, to fine-tune a unified model jointly on all locales, helps to overcome this problem.

In the area of multi-locale query classification, \citet{lin2020light} develop a shard selection method for e-commerce product search. They conduct experiments under high-resource conditions, thus not addressing the low-resource issue in the smaller stores. Close to our work is \cite{luo2021q2pt}, proposing a BERT-based query classification model, which has a separate product type classification layer for each locale. The unified backbone allows to reduce storage requirements, however, training the classification layer for each locale separately reduces the model's ability to transfer knowledge.
We use the architecture from \citet{luo2021q2pt} as one of the baselines for our experiments.

Similar to this study, \citet{roitero2020leveraging} compare the performance of global and locale-specific models in cross-market music recommendation domain, showing that market knowledge plays an important role.

%Finally, in machine translation domain, the language id is prepended to the input text to condition on the source/target language of translation (\citet{wicks2022effects} give an extensive overview of the related models). We borrow this approach for our experiments, benchmarking the Q2PT model with the locale id prepended to the input query.

%\annant{add section on transfer learning?}

\section{Methodology}
\label{sec:method}

We study the task of predicting product type intent for a given e-commerce search query (Q2PT) in a multi-locale setting. This is a multi-label classification task: for an input query in each locale, the Q2PT model needs to return all product types associated with the query. In the multi-locale setup, each store has an independent catalogue of items, potentially a different language and different volumes of traffic.

In our study we propose an alternative to an existing method of training Q2PT models for each locale separately \cite{luo2021q2pt}, replacing it with a unified architecture, which shares model parameters and training data across all stores. 

We use BERT-based encoder \cite{devlin2018bert} to create the representations of the input queries, following current research in query classification~\cite{luo2021q2pt, zhu2022enhanced, zhu2023hcl4qc, qiu2022pre}. The encoder is followed by a fully-connected layer with sigmoid activation, applied to the [CLS] token.

Building on this base architecture, we compare three variants:
\vspace{-0.1cm}
\begin{itemize}
    \item \textbf{non-unified (\noncons)} \cite{luo2021q2pt} - which has a common DistilBert backbone but separate classifiers for all locales. %(see Figure \ref{fig:model_architecture}, left panel). 
    Each locale-specific classifier is trained only on the data from this specific store.
    \item \textbf{unified locale-agnostic (\consnomp)} - in this model both DistilBert encoder and classifier are shared across all locales and trained on the mixture of the global data.
    \item \textbf{unified locale-aware (\consmp)} - this model is similar tho the locale-agnostic version, however, it conditions the product type prediction on the locale-id. To achieve it, we prepend the locale-id token to the input keywords, separated with a special [SEP] token.% (see Figure \ref{fig:model_architecture}, right pane).
\end{itemize}

Both unified model variants have their merits and drawbacks. \consnomp solves the problem of data shortage in low-resource stores, but does not account for local specifics in different locales. Specifically, it can lead to biases in smaller locales, transferred from the bigger ones. \consmp overcomes this problem by encoding the input locale information alongside with the query. This effectively allows the model to produce different product type distributions for each~locale.

On the other hand, locale-agnostic model has its advantages over locale-aware one: \consnomp model is more practical for the cases of cold-start launches of new stores, with no prior training data. In this scenario \consmp needs to be retained to learn about new locale-id, while \consnomp can be readily used out of the box.

\section{Experimental Setup}
\label{sec:experiments}

\subsection{Datasets}

We use an experimental data set sampled from real data, covering 1414 product types and 20 locales: (\textbf{US} - United States, \textbf{DE} - Germany, \textbf{UK} - United Kingdom, \textbf{JP} - Japan, \textbf{IN} - India, \textbf{IT} - Italy, \textbf{CA} - Canada, \textbf{FR} - France, \textbf{ES} - Spain, \textbf{MX} - Mexico, \textbf{BR} - Brazil, \textbf{AE} - United Arab Emirates, \textbf{AU} - Australia, \textbf{SA} - Saudi Arabia, \textbf{EG} - Egypt, \textbf{NL} - Netherlands, \textbf{TR} - Turkey, \textbf{SE} - Sweden, \textbf{SG} - Singapore, \textbf{PL} - Poland). To the best of our knowledge, there are no existing studies benchmarking query classification models across this large number of locales.
%We filter the dataset to only include \textit{tail} queries: the ones that constitute the lowest 30\% of all queries, sorted by frequency. Tail queries are extremely challenging, especially in the low-resource locales, for both product matching and query classification, thus allowing to perform a stress-test on the benchmark models.

\subsection{Training data} 
\label{sec:traindata}

The data for training Q2PT models is created by aggregating fully anonymized customer click-through behavior, following previous research \cite{liu2019system, zhang2021modeling, lin2020light}. The product type for the user search query is derived as the majority product type of items, that the user clicked following that query. Formally, we define the probability of the query $q$ belonging to the product type $p$ as follows:
\begin{equation}
    P(q,p) = \frac{num\_clicks(q,item | item \in p)}{num\_clicks(q, item)}
    \label{eq:1}
\end{equation}

After that we select all <\textit{query, product type}> pairs that have $P(q,p) > 0.5$.

%In total the training data amounted to 9.7B instances. The number of samples per locale are presented in Table~\ref{tab:human_exp} and Figure~\ref{fig:per_locale}, illustrating the immense discrepancy of the data distribution among the stores: the high-resource stores take 94\% share of the chart. 

\begin{comment}
    
\begin{figure}[ht!]
        \centering
        \includegraphics[width=0.95\linewidth]{figures/new_chart.jpg}
        \caption{Distribution of queries per locale.}
    \label{fig:per_locale}
    %\vspace{-1cm}
\end{figure}

\end{comment}

The collected data has immense discrepancy in sample distribution among the locales: the large stores take over 90\% share of the dataset. We split the locales into 2 buckets: \textit{High-Resource} (Hi-Re) locales, including US, DE, UK, JP, IN, IT, CA, FR, ES stores, and \textit{Low-Resource} (Lo-Re) locales, including MX, BR, AE, AU, SA, EG, NL, TR, SE, SG, PL stores, based on the number of training samples. In our experiments we compare the models' results per bucket, to assess the effects of unified and disjoint training for these two groups.

We use 100s of millions samples for model training and 10s of millions for validation and hyperparameter selection.%\anna{changed as proposed}

\begin{table*}[]
\centering
    \adjustbox{max width=0.85\textwidth}{
\begin{tabular}{cccccccccccc|c}
 & \textbf{MX} & \textbf{BR} & \textbf{AE} & \textbf{AU} & \textbf{SA} & \textbf{EG} & \textbf{NL} & \textbf{TR} & \textbf{SE} & \textbf{SG} & \textbf{PL} & \textbf{Lo-Re} \\ \hline
%\# samples & 143 & 70.2 & 52.4 & 51.6 & 50.1 & 41.9 & 38.7 & 22.3 & 21.4 & 11.8 & 9.7 & 540.5 \\ \hdashline
\textbf{\noncons} & 0.91 & 0.81 & 0.9 & 0.84 & 0.83 & 0.72 & 0.66 & 0.87 & 0.53 & 0.91 & 0.54 & 0.77 \\
\textbf{\consnomp} & 0.92 & 0.78 & 0.9 & 0.85 & 0.87 & 0.77 & 0.69 & 0.9 & 0.55 & 0.91 & 0.59 & 0.79 (+2\%)\\
\textbf{\consmp} & 0.92 & 0.82 & 0.91 & 0.87 & 0.87 & 0.76 & 0.70 & 0.90 & 0.58 & 0.92 & 0.60 & 0.80 (+3\%)\\

 \hline \hline
\end{tabular}
} 
\\ \vspace{2mm}
    \adjustbox{max width=\textwidth}{
\begin{tabular}{cccccccccc|c|c}

 & \textbf{US} & \textbf{DE} & \textbf{UK} & \textbf{JP} & \textbf{IN} & \textbf{IT} & \textbf{CA} & \textbf{FR} & \textbf{ES} & \textbf{Hi-Re} & \textbf{WW} \\ \hline
%\# samples & 4667.2 & 912.8 & 730.8 & 615.2 & 562 & 453.6 & 426.5 & 374.8 & 315.9 & 9058.8 & 9599.5 \\ \hdashline
\textbf{\noncons} & 0.79 & 0.9 & 0.88 & 0.76 & 0.85 & 0.93 & 0.92 & 0.89 & 0.92 & 0.87 & 0.82 \\
\textbf{\consnomp} & 0.81 & 0.91 & 0.9 & 0.79 & 0.83 & 0.95 & 0.93 & 0.91 & 0.93 & 0.89 (+2\%) & 0.83 (+1\%)\\
\textbf{\consmp} & 0.80 & 0.91 & 0.90 & 0.80 & 0.86 & 0.95 & 0.92 & 0.91 & 0.93 & 0.89 (+2\%) & 0.84 (+2\%)\\
         
\end{tabular}
}
\vspace{3mm}
\caption{Results for recall at 0.8 precision, comparing the baseline (\noncons) and the unified methods (\consnomp, \consmp), in Low-Resource (top) and High-Resource (bottom) locales on the human-labeled evaluation set.}
    \label{tab:human_exp}
\end{table*}

\subsection{Evaluation Data}

As the customer click-through data is noisy and prone to trends and seasonality, we cannot rely on it for accurate model performance evaluation. Instead, we create two separate evaluation datasets: human-annotated and automatically weakly~labeled.

\paragraphHd{Human-labeled data.} We recruited professional annotators to label search queries with all applicable product types; for each locale we got around 1k human-labeled queries. The human-annotated dataset is high-quality, however, it mostly consists of queries associated with popular PTs. Thus, this dataset has a very low coverage of the product type label space: on average only around 600 product types (42\% of the whole list) are included for each locale, moreover, on average only 22 PTs are associated with at least 20 labeled samples. 

As an alternative, to be able to conduct a more fine-grained per-PT anaysis, we created a large-scale automatically labeled dataset.

\paragraphHd{Automatically labeled data.} To create this dataset we leverage relevance labels for the <\textit{query, item}> pairs, obtained from a pre-trained classifier. For each query we collect the categories of all items that are predicted relevant to it. The query label is then selected as a majority category from relevant items. 

The resulting dataset consists of 2.7M labels in total, with an average of 120k labels per locale. To check the correctness of this labelling, we manually inspected 200 queries from different locales, and verified that this labeling method achieves almost 90\% accuracy. Importantly, this method allowed us to collect a sufficient number of evaluation samples for all 1414 product types. 

Nevertheless, the automated approach has the following drawbacks: i) by taking the majority product type label, we make an assumption that the query can be associated with only one product type. This is a valid assumption in most cases, but there are some exceptions (e.g. a query `\textit{sun protection}' can mean a \textit{cream}, \textit{apparel} or \textit{glasses}); 
ii) the quality of the resulting annotations depend on the quality of the relevance classifier; iii) the distribution of the queries and product types in this dataset does not match the one in the training~data.

Taking into account these drawbacks of the automated approach, we opt to use human-annotated queries as our main evaluation data, and use the automatically-labeled one for an additional in-depth per-PT analysis.

\subsection{Model Implementation and Training}

We used a multilingual pretrained DistilBert~\citep{sanh2019distilbert} checkpoint, which was fine-tuned on e-commerce queries; we further fine-tuned the model on the Q2PT task. We used Adam optimizer and trained the model with 8e-5 learning rate, 0.001 dropout and $2^{11}$ batch size until convergence, using binary cross-entropy loss. The hyperparameters were chosen through grid search on the validation split. 
%, which is sample-weighted with the query-product type click probabilities from equation 1, to incorporate the uncertainty of the click-trough data aggregation; 
%for inference we select the product type with the highest score from the classification layer.
%\annant{we select the product types which confidence score surpass the confidence threshold determined on the validation dataset (0.4)}

\subsection{Evaluation Metrics}

We report recall at 0.8 precision: the fixed high-precision setting is a standard in evaluating customer-facing applications, such as e-commerce sites. Given the importance of Q2PT signal for the downstream search and recommendation components, the query classification models have to meet high precision bar.

\section{Experiment Results}
\label{sec:offline}

% not to lose it
% https://quip-amazon.com/o0wlAuGvleao/Consolidated-Labels-Recall-Evaluation-Missing-PTs-Deep-Dive
%\annant{do we need to add sth about the language differences: this is why sg performs still well}

%\annant{maybe do AUC}

\paragraphHd{Results on the human-labeled data.}
In Table \ref{tab:human_exp} we report recall at 0.8 precision for all 20 locales. Additionally, we aggregate the results separately for established High-Resource (Hi-Re) and emerging Low-Resource (Lo-Re) stores, and worldwide (WW).%\anna{adjusted the description of this section not mentioning the sizes of locales}

We observe that both variants of the unified model outperform the non-unified one for all locales, with the total increase of 2\% recall worldwide. Notably, the most benefiting locales are small ones, e.g. PL (+6\%) and SE (+5\%). It shows that the unified models can efficiently transfer knowledge from high-resource to the low-resource locales. Importantly, we notice that even on the Hi-Re locales a generalist \consmp model performs slightly better than a specialist \consmp model.

\begin{figure*}[htp]
    \centering
    \begin{minipage}{.5\textwidth}
        \centering
        \includegraphics[width=0.9\linewidth]{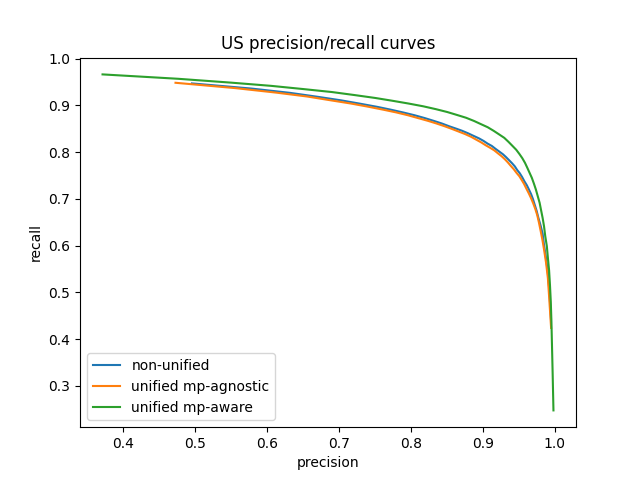}
    \end{minipage}%
    \begin{minipage}{0.5\textwidth}
        \centering
        \includegraphics[width=0.9\linewidth]{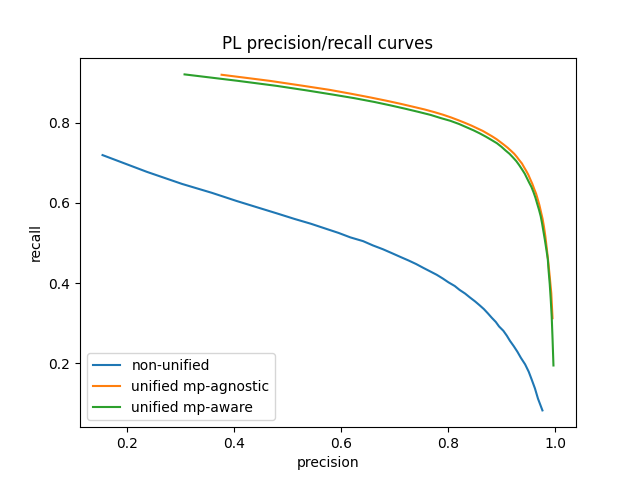}
    \end{minipage}
    \caption{Precision-recall curves in US (left) and PL (right) locales.}
    \label{fig:curves}
\end{figure*}

Between the two variants of a unified model, \consmp has slightly more pronounced gains over \consnomp. This illustrates that the task of product type prediction is not locale-invariant: conditioning on the locale information helps the model to distinguish locale-specific peculiarities (we discuss it further in Section \ref{sec:differences}). 

To further assess the models on low- and high-resource locales we plot precision-recall curves for one of the high-resource (US) and one of the small-resource (PL) locales in Figure \ref{fig:curves}. For US, \consmp slightly dominates for the high precision values, thanks to the both more diverse training data from the other locales and preserved information about the current locale. In PL, there is a significant gap between \noncons and consolidated models, because the latter were trained on 1000 times more data.

\vspace{0.5cm}

Additionally, we experimented with a completely disjoint model architecture: for each locale all model parts are shared among the locales, and they are trained and stored separately. We found that this variation performs on par with \consmp and has 2\% performance increase compared no \noncons. Given that this model takes 7 times longer to train (assuming sequential training on all locales) and 20 times greater memory requirements, without significant performance improvements, we did not consider this model for our analyses.

\paragraphHd{Results on the automatically-labeled data.} We compute recall at 0.8 precision on the automatically-labeled evaluation set, and present the aggregated results in Table \ref{tab:auto_exp}. On this dataset the gap between \noncons and unified models considerably increases, which can be attributed to the automated data having more long-tail product types, which are challenging to predict, and which have extremely scarce training data in low-resource locales. At the same time, human data is largely composed of the most popular product types, on which the performance of all models is on par.

In this dataset, however, the trend between both unified models changes, with \consnomp having slightly better performance. After examining the performance of the models per locale, we found that \consnomp has superior results in all but two of the biggest locales: US and UK (in these locales \consnomp performance is 1\% worse than \consmp). It shows that conditioning on locale-id helps the model to develop some locale-specific knowledge, which is abundant in high-resource locales, and thus improves the performance of \consmp.

\paragraphHd{Qualitative analysis.} We also note that the knowledge transfer can have negative effect on the Q2PT predictions in some cases. To validate it, we checked the evaluation queries for which the \noncons model made a correct prediction, while a unified model (we use \consmp in this study) made an error. For instance, the query `\textit{roma}' should yield different product types depending on the store: it should return \textit{laundry\_detergent} for MX (local brand), \textit{personal\_fragrance} for DE and other European stores (name of perfume popular in Europe), and for the remaining locales the correct prediction will be a \textit{book}. We notice that in SG \consmp model, biased by the data from the other locales, predicts \textit{laundry\_detergent} for the query `\textit{roma}', while \noncons correctly predicts \textit{book}.

\begin{table}[]
\begin{tabular}{lccc}
                  & \textbf{Lo-Re} & \textbf{Hi-Re} & \textbf{WW} \\ \hline
\textbf{\noncons}  & 0.64           & 0.80           & 0.71        \\
\textbf{\consnomp} & 0.83 (+17\%)          & 0.87 (+7\%)          & 0.85 (+14\%)       \\
\textbf{\consmp}     & 0.81 (+15\%)          & 0.86 (+6\%)          & 0.83 (+12\%)      
\end{tabular}
\vspace{1em}
\caption{Results for recall at 0.8 precision, comparing the baseline (\noncons) and the unified methods (\consnomp, \consmp), on the automatically labeled evaluation set.}
    \vspace{-0.3cm}
\label{tab:auto_exp}
\end{table}

Another interesting dimension of analysis is the differences between the errors of \consnomp and \consmp on the automatically-labeled dataset. Given that US and UK are the only locales where \consmp outperforms the locale-agnostic model, we investigated the US queries where \consnomp fails, while \consmp is correct. One example query is `\textit{boys drawers}': while `drawers' is a common name for underwear in Asia, in the US this term is not used, and the locale-aware model predicts the correct product type as \textit{dresser}. At the same time, \consnomp model, which does not have access to the locale information, erroneously predicts PT \textit{underpants}. 

Another example is query `\textit{vanity side shelf with drawers}': in the US the term `vanity', among other things, refers to furniture, while in British English this is uncommon. Therefore, the products categorized as \textit{makeup\_vanity} in US, have \textit{table} product type in other stores. This influences the training data for the models, which is based on the user clicks on the product categories, as described in Section \ref{sec:traindata}. Therefore, in this case for the query `\textit{vanity side shelf with drawers}' in US locale \consmp model predicts the product type \textit{makeup\_vanity}, while \consnomp predicts \textit{table}.

Finally, we note that such cases of meaningful discrepancies are rare (less than 1\% of the automatic evaluation set), as mostly the differences in the models' predictions arise from one of the models refraining from prediction.

\begin{table}[]
\begin{tabular}{ccccccc}
            & \multicolumn{2}{c}{\textbf{WW}}                         & \multicolumn{2}{c}{\textbf{Lo-Re}}                      & \multicolumn{2}{c}{\textbf{Hi-Re}} \\
            & \textbf{\noncons} & \multicolumn{1}{c|}{\textbf{\consmp}} & \textbf{\noncons} & \multicolumn{1}{c|}{\textbf{\consmp}} & \textbf{\noncons} & \textbf{\consmp} \\ \hline
correlation & 0.2              & \multicolumn{1}{c|}{0.15}            & 0.22             & \multicolumn{1}{c|}{0.15}            & 0.16             & 0.14            \\ \hdashline
head        & 0.85             & \multicolumn{1}{c|}{0.9}             & 0.83             & \multicolumn{1}{c|}{0.89}            & 0.87             & 0.91            \\
torso       & 0.79             & \multicolumn{1}{c|}{0.85}            & 0.77             & \multicolumn{1}{c|}{0.84}            & 0.82             & 0.87            \\
tail        & 0.7              & \multicolumn{1}{c|}{0.8}             & 0.65             & \multicolumn{1}{c|}{0.78}            & 0.75             & 0.83           
\end{tabular}
\caption{Pearson correlation of the number of samples and accuracy per-PT (top line), accuracy per product type on the head/torso/tail PT splits.}
\label{tab:auto_exp}
\vspace{-0.8cm}
\end{table}

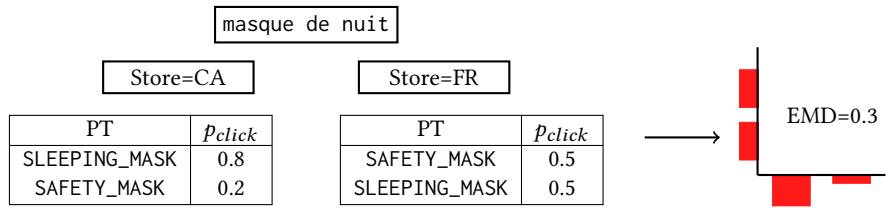
\begin{figure*}[h!]
    \centering
    \begin{tikzpicture} %[scale=0.7, every node/.style={scale=0.7}]
        \node(kw) [draw,thick,minimum width=2cm] at (-1,0) {\texttt{masque de nuit}};
        \node(sa) [draw,thick,minimum width=2cm] at (-2.7,-0.7) {Store=CA};
        \node(sb) [draw,thick,minimum width=2cm] at (0.7,-0.7) {Store=FR};
        
        \node(ta) at (-3.2,-1.8) {
        \begin{tabular}{|c|c|}
            \hline
            PT & $p_{click}$\\
            \hline
            \texttt{SLEEPING\_MASK} & 0.8 \\
            \texttt{SAFETY\_MASK} & 0.2 \\
            \hline
        \end{tabular}
        };
        
        \node(tb) at (1.2,-1.8) {
        \begin{tabular}{|c|c|}
            \hline
            PT & $p_{click}$\\
            \hline
            \texttt{SAFETY\_MASK} & 0.5 \\
            \texttt{SLEEPING\_MASK} & 0.5 \\
            \hline
        \end{tabular}
        };
        
        \draw [->, thick] (3.5, -1.5) -- (4.5, -1.5);
        
        % Store CA
        \draw [fill,color=red!90,draw] (5.2,-2.01) -- (5.7,-2.01) -- (5.7,-2.41) -- (5.2,-2.41) -- (5.2,-2.01);
        \draw [fill,color=red!90,draw] (6,-2.01) -- (6.5,-2.01) -- (6.5,-2.11) -- (6,-2.11) -- (6,-2.01);
        
        % Store FR
        \draw [fill,color=red!90,draw] (5.01,-1.8) -- (5.01,-1.3) -- (4.76, -1.3) -- (4.76, -1.8) -- (5.01,-1.8);
        \draw [fill,color=red!90,draw] (5.01,-1.1) -- (5.01,-0.6) -- (4.76, -0.6) -- (4.76, -1.1) -- (5.01,-1.1);
        
        % XAxis
        \draw [thick] (5, -2) -- (6.7, -2);
        % YAxis
        \draw [thick] (5, -2) -- (5, -0.3);
        
        \node at (6,-1.2) {EMD=$0.3$};
        
        %\node at (8, -2) {\input(kde.pgf}};
    \end{tikzpicture}
    \caption{The query ``\textit{masque de nuit}'' exists in both training sets for CA and FR. In CA, the most clicked PT is \texttt{SLEEPING\_MASK} while in FR, \texttt{SAFETY\_MASK} and \texttt{SLEEPING\_MASK} have equal rates, which results in EMD of 0.3}
    \label{fig:emd}
\end{figure*}

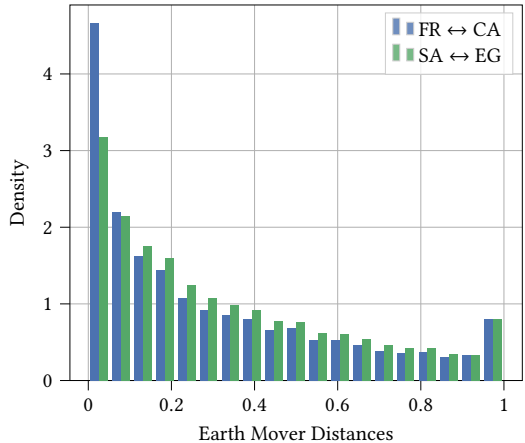
\begin{figure}[h]
    \centering
    \adjustbox{width=0.4\textwidth}{
    % This file was created with tikzplotlib v0.10.1.
\begin{tikzpicture}

\definecolor{darkgray176}{RGB}{176,176,176}
\definecolor{lightgray204}{RGB}{204,204,204}
\definecolor{mediumseagreen85168104}{RGB}{85,168,104}
\definecolor{steelblue76114176}{RGB}{76,114,176}

\begin{axis}[
legend cell align={left},
legend style={fill opacity=0.8, draw opacity=1, text opacity=1, draw=lightgray204},
tick align=outside,
tick pos=left,
x grid style={darkgray176},
xlabel={Earth Mover Distances},
xmajorgrids,
xmin=-0.0442105263157895, xmax=1.04421052631579,
xtick style={color=black},
y grid style={darkgray176},
ylabel={Density},
ymajorgrids,
ymin=0, ymax=4.89986348071665,
ytick style={color=black}
]
\draw[draw=none,fill=steelblue76114176] (axis cs:0.00526315789473684,0) rectangle (axis cs:0.0263157894736842,4.66653664830157);
\addlegendimage{ybar,ybar legend,draw=none,fill=steelblue76114176}
\addlegendentry{FR $\leftrightarrow$ CA}

\draw[draw=none,fill=steelblue76114176] (axis cs:0.0578947368421052,0) rectangle (axis cs:0.0789473684210526,2.19583616185786);
\draw[draw=none,fill=steelblue76114176] (axis cs:0.110526315789474,0) rectangle (axis cs:0.131578947368421,1.62319027562643);
\draw[draw=none,fill=steelblue76114176] (axis cs:0.163157894736842,0) rectangle (axis cs:0.184210526315789,1.44533940419084);
\draw[draw=none,fill=steelblue76114176] (axis cs:0.215789473684211,0) rectangle (axis cs:0.236842105263158,1.08246889873195);
\draw[draw=none,fill=steelblue76114176] (axis cs:0.268421052631579,0) rectangle (axis cs:0.289473684210526,0.92260406782458);
\draw[draw=none,fill=steelblue76114176] (axis cs:0.321052631578947,0) rectangle (axis cs:0.342105263157895,0.858940325320952);
\draw[draw=none,fill=steelblue76114176] (axis cs:0.373684210526316,0) rectangle (axis cs:0.394736842105263,0.808288670774747);
\draw[draw=none,fill=steelblue76114176] (axis cs:0.426315789473684,0) rectangle (axis cs:0.447368421052632,0.661535691916483);
\draw[draw=none,fill=steelblue76114176] (axis cs:0.478947368421053,0) rectangle (axis cs:0.5,0.679835276732839);
\draw[draw=none,fill=steelblue76114176] (axis cs:0.531578947368421,0) rectangle (axis cs:0.552631578947368,0.528820945958616);
\draw[draw=none,fill=steelblue76114176] (axis cs:0.584210526315789,0) rectangle (axis cs:0.605263157894737,0.527623776858482);
\draw[draw=none,fill=steelblue76114176] (axis cs:0.636842105263158,0) rectangle (axis cs:0.657894736842105,0.4587723015876);
\draw[draw=none,fill=steelblue76114176] (axis cs:0.689473684210526,0) rectangle (axis cs:0.710526315789474,0.382780567755195);
\draw[draw=none,fill=steelblue76114176] (axis cs:0.742105263157895,0) rectangle (axis cs:0.763157894736842,0.354020005325752);
\draw[draw=none,fill=steelblue76114176] (axis cs:0.794736842105263,0) rectangle (axis cs:0.81578947368421,0.370538088505002);
\draw[draw=none,fill=steelblue76114176] (axis cs:0.847368421052631,0) rectangle (axis cs:0.868421052631579,0.302470473954446);
\draw[draw=none,fill=steelblue76114176] (axis cs:0.9,0) rectangle (axis cs:0.921052631578947,0.32873693409194);
\draw[draw=none,fill=steelblue76114176] (axis cs:0.952631578947368,0) rectangle (axis cs:0.973684210526316,0.801661484684711);
\draw[draw=none,fill=mediumseagreen85168104] (axis cs:0.0263157894736842,0) rectangle (axis cs:0.0473684210526316,3.18282793160111);
\addlegendimage{ybar,ybar legend,draw=none,fill=mediumseagreen85168104}
\addlegendentry{SA $\leftrightarrow$ EG}

\draw[draw=none,fill=mediumseagreen85168104] (axis cs:0.0789473684210526,0) rectangle (axis cs:0.1,2.14266405693551);
\draw[draw=none,fill=mediumseagreen85168104] (axis cs:0.131578947368421,0) rectangle (axis cs:0.152631578947368,1.75495353519492);
\draw[draw=none,fill=mediumseagreen85168104] (axis cs:0.184210526315789,0) rectangle (axis cs:0.205263157894737,1.59762095711329);
\draw[draw=none,fill=mediumseagreen85168104] (axis cs:0.236842105263158,0) rectangle (axis cs:0.257894736842105,1.24656631539984);
\draw[draw=none,fill=mediumseagreen85168104] (axis cs:0.289473684210526,0) rectangle (axis cs:0.310526315789474,1.07921730498522);
\draw[draw=none,fill=mediumseagreen85168104] (axis cs:0.342105263157895,0) rectangle (axis cs:0.363157894736842,0.985979238055713);
\draw[draw=none,fill=mediumseagreen85168104] (axis cs:0.394736842105263,0) rectangle (axis cs:0.415789473684211,0.916929789632488);
\draw[draw=none,fill=mediumseagreen85168104] (axis cs:0.447368421052632,0) rectangle (axis cs:0.468421052631579,0.774888254527304);
\draw[draw=none,fill=mediumseagreen85168104] (axis cs:0.5,0) rectangle (axis cs:0.521052631578947,0.769507086093087);
\draw[draw=none,fill=mediumseagreen85168104] (axis cs:0.552631578947368,0) rectangle (axis cs:0.573684210526316,0.619367158888882);
\draw[draw=none,fill=mediumseagreen85168104] (axis cs:0.605263157894737,0) rectangle (axis cs:0.626315789473684,0.604662183761715);
\draw[draw=none,fill=mediumseagreen85168104] (axis cs:0.657894736842105,0) rectangle (axis cs:0.678947368421053,0.541260298249649);
\draw[draw=none,fill=mediumseagreen85168104] (axis cs:0.710526315789474,0) rectangle (axis cs:0.731578947368421,0.457346038013089);
\draw[draw=none,fill=mediumseagreen85168104] (axis cs:0.763157894736842,0) rectangle (axis cs:0.784210526315789,0.425005748512397);
\draw[draw=none,fill=mediumseagreen85168104] (axis cs:0.815789473684211,0) rectangle (axis cs:0.836842105263158,0.420423763509005);
\draw[draw=none,fill=mediumseagreen85168104] (axis cs:0.868421052631579,0) rectangle (axis cs:0.889473684210526,0.340771814903509);
\draw[draw=none,fill=mediumseagreen85168104] (axis cs:0.921052631578947,0) rectangle (axis cs:0.942105263157895,0.336402945481669);
\draw[draw=none,fill=mediumseagreen85168104] (axis cs:0.973684210526316,0) rectangle (axis cs:0.994736842105263,0.803605579141592);
\end{axis}

\end{tikzpicture}
    }
    \caption{Density of EMD distances for queries in the intersection of FR $\leftrightarrow$ CA and SA $\leftrightarrow$ EG pairs.}
    \label{fig:emd_kde}
\vspace{-0.5cm}
\end{figure}

\subsection{Analysis on the Product Type Level} 

In this analysis we aim to evaluate the models on different product type groups, based on their frequency in the training data. As shown in Figure \ref{fig:PT_percent}, the distribution of product types is very skewed, and thus it is desirable to assess the models' capabilities to correctly infer long-tail product types. 

For this analysis we split the list of PTs into 3 parts: \textit{head} (very frequent PTs, e.g. \textit{book}), \textit{torso} (average frequency, e.g. \textit{wall ornament}) and \textit{tail} (niche PTs, e.g. \textit{mounted storage system kit}), so that each part has 1/3 of query mass of the whole dataset. As a result we got 48 head, 198 torso and 1168 tail product types.

For this experiment we use automatically-labeled data, because it evenly covers all product types. We compare the performance of non-unified model against unified locale-aware model, as it has shown superior performance to the \consnomp on the human-labeled benchmark. We compute per-PT accuracy, which is defined as the number of correctly predicted occurrences of a product type, over all occurrences of this product type in the evaluation set. Additionally, we compute Pearson correlation of the PT frequency and PT accuracy, to see how much the models overfit on the high-frequency product types. 

The results of this analysis are shown in Table \ref{tab:auto_exp}, aggregated for low- and high-resource locales and worldwide.

Unsurprisingly, both models perform better on the more frequent groups of product types, however, the accuracy drop from head to tail PT group is more prominent for the non-unified model. Additionally, we noted that the unified model dramatically reduces the number of product types, which don't meet a high accuracy bar of at least 0.5. Non-unified model has 105 such low-accuracy PTs, compared to 27 for a unified model. Most of those 27 PTs are related to digital content, which are easy to confuse to each other (e.g., \textit{music track} and \textit{music album}).

Consistently with the previous results, the correlation of accuracy and PT frequency is greater in non-unified model. One interesting observation is that the correlation difference is small on Hi-Re locales and is more pronounced on the Lo-Re stores. Additionally, we compared correlations in the biggest locale (US) and the smallest (PL). The correlation in US was similarly low for both \noncons and \consmp (0.12 and 0.9 resp.), however, in PL the correlation of \noncons accuracy is extremely high (0.23) and it is getting smoothed with \consmp (0.11).

\section{Discussion}

\subsection{Analysis of Locale Differences}
\label{sec:differences}

In our experiments we observe that the same search query might have different meanings depending on the locale, and thus yield different product type distributions. We conduct additional analyses to quantify how frequent this phenomenon occurs.

We focus on two pairs of locales that share the same language: FR~$\leftrightarrow$~CA (French) and EG $\leftrightarrow$ SA (Arabic); for each pair of stores, we consider the queries that exist in the training sets in both markets and measure the difference between per-PT click distributions (from Equation \ref{eq:1}). To this end, we compute the classic Earth Moving Distance (EMD)~\citep{monge1781memoire, bonneel2011displacement}. In Figure~\ref{fig:emd} we illustrate one example of such computation.

We depict the histograms of EMD distributions for each locale pair in Figure~\ref{fig:emd_kde}: the smaller the EMD, the more similar the locale per-PT distributions are. We observe an expected peak near zero, because largely the customers in different stores should have the same PT in mind in their search. However, the amount of dissimilar modes is significant, ranging from small differences to completely different distributions (EMD near 1).

\subsection{Categorization of Local Differences}

\label{sec:typo}

\begin{figure}[ht!]
\centering
\begin{subfigure}{0.48\textwidth}
    \includegraphics[width=\textwidth]{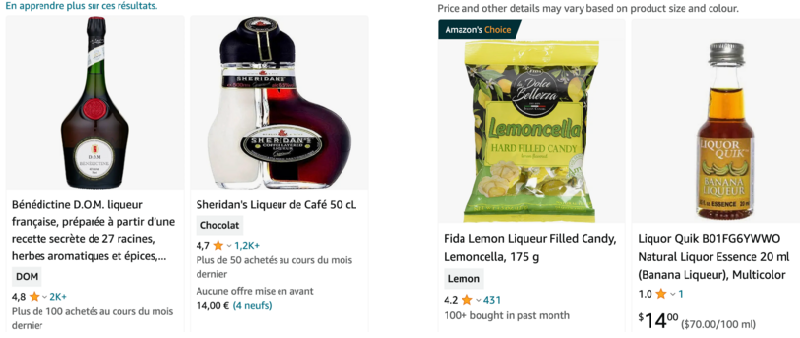}
    \caption{\texttt{liquere} products}
    \vspace{0.5cm}
    \label{fig:local_diff_liquere}
\end{subfigure}
\begin{subfigure}{0.48\textwidth}
    \includegraphics[width=\textwidth]{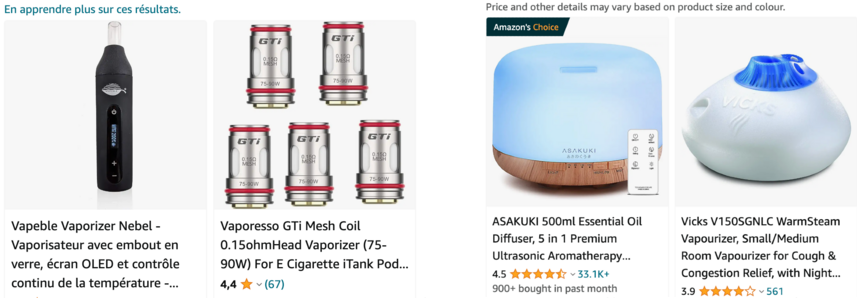}
    \caption{\texttt{vaporizer} products}
    \vspace{0.5cm}
    \label{fig:local_diff_vaporizer}
\end{subfigure}
\hfill
\begin{subfigure}{0.48\textwidth}
    \includegraphics[width=\textwidth]{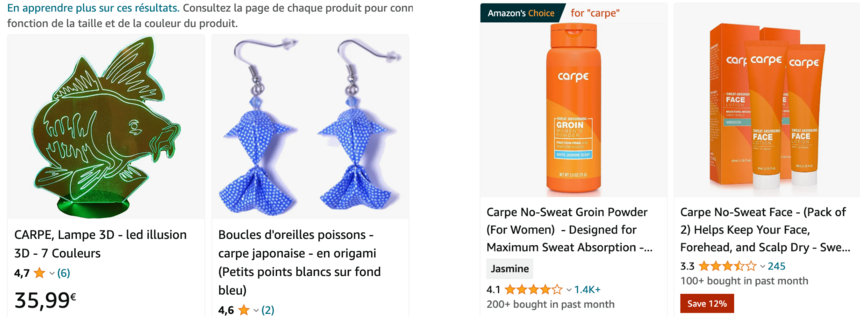}
    \caption{\texttt{carpe} products}
    \label{fig:local_diff_carpe}
\end{subfigure}
        
\caption{Examples of product type discrepancies between FR (left 2 products) and CA (right 2 products) stores.}
\label{fig:local_diff}
\vspace{-0.5cm}
\end{figure}

During analysis of query differences, we found multiple cases where the product type distributions significantly vary across locales. We attribute those discrepancies to one of the following cases:

\textbf{1. Dialectal differences}: despite the same language, the query has a different meaning to the users in different stores. One particular example is the word `\textit{liqueur}', which means an alcoholic drink in France, but a non-alcoholic drink or syrup in Canada (see Figure~\ref{fig:local_diff_liquere}). Another example is `\textit{vaporizer}', which means a smoking gadget in France and an air humidifier appliance in Canada (depicted in Figure~\ref{fig:local_diff_vaporizer}). These discrepancies can be explained by cultural and historical factors, e.g. language drift from the neighbouring countries. Although such cases are pretty rare in the data, they can have a profound impact on the user experience, especially in the cases of products under legal regulations.

\textbf{2. Selection differences}: the query has different meaning with respect to the product type, that is caused by a mismatch in the selection of products offered in corresponding markets. For example, a query ``carpe'' that means ``a carp'' in French leads to fishing accessories in FR store. However, in Canada it is mapped to a popular cosmetic brand marketed under the same name (see Figure~\ref{fig:local_diff_carpe}). %PTs with a limited offer in particular store also belong to this category.

\textbf{3. Noisy differences}: in many cases we observe a large discrepancy for a given query as measured by EMD between PT click distributions. However, this metric is not always conclusive since EMD doesn't account for uncertainties in click probabilities from Equation \ref{eq:1}. As a toy example, a PT with 5 clicks and 10 impressions should be treated differently than PT with 10K clicks and 20K impressions, even though the probability $p$ will be equal to $0.5$ in both cases.

\section{Conclusion and Future Work}

In this study we investigate the task of e-commerce query product type prediction in a multi-locale setup and propose a transfer learning solution to augment Q2PT predictions in low-resource stores to achieve global parity. 

The evidence from our offline experiments and user studies shows that the unified Q2PT model, sharing the parameters and training data across all locales, outperforms the non-unified model, on both low- and high-resource locales, additionally decreasing infrastructure requirements. We compare locale-agnostic and locale-aware variants of the unified model, showing that it is important to capture store-specific characteristics by conditioning the prediction on the locale-id. The findings from our work will be useful for practitioners developing multi-locale query classification~models.

We identify the following avenues for our future work. We plan to investigate the impact of language variation on the Q2PT task, both in terms of the language of the input query as well as the model's pre-training data. Additionally, we plan to conduct similar analyses for other related query understanding models, such as brand classifiers.

\bibliographystyle{ACM-Reference-Format}
\balance
\bibliography{bibliography}

\appendix

\end{document}